\documentclass[conference]{IEEEtran}
\IEEEoverridecommandlockouts

\usepackage{epsfig}
\usepackage{epsf}
\usepackage{psfrag}
\usepackage{cite}
\usepackage[cmex10]{amsmath}
\usepackage{amsfonts}
\usepackage{amssymb}
\usepackage{graphicx}
\usepackage{multirow}
\usepackage{slashbox}
\usepackage{stfloats}

\newtheorem{theorem}{Theorem}
\newtheorem{lemma}{Lemma}

\newtheorem{example}{Example}
\newtheorem{remark}{Remark}

\newcommand{\qed}{\hfill \rule{2mm}{2mm}}

\newcommand{\ul}[1]{\underline{#1}}

\newcommand{\set}[1]{{\cal #1}}

\begin{document}
\author{
\IEEEauthorblockN{Tobias Lutz}
\IEEEauthorblockA{Inst. for Communications Engineering\\
Technische Universit\"{a}t M\"{u}nchen\\
Munich, 80290 Germany\\
tobi.lutz@tum.de}
\and
\IEEEauthorblockN{Gerhard Kramer}
\IEEEauthorblockA{Department of Electrical Engineering\\
University of Southern California\\
Los Angeles, CA 90089 USA\\
gkramer@usc.edu}
\and
\IEEEauthorblockN{Christoph Hausl}
\IEEEauthorblockA{Inst. for Communications Engineering\\
Technische Universit\"{a}t M\"{u}nchen\\
Munich, 80290 Germany\\
christoph.hausl@tum.de}
}
\title{Capacity for Half-Duplex Line Networks with Two Sources}
\maketitle

\begin{abstract}
The focus is on noise-free half-duplex line networks with two sources where the first node and either the second node or the second-last node in the cascade act as sources. In both cases, we establish the capacity region of rates at which both sources can transmit independent information to a common sink. The achievability scheme presented for the first case is constructive while the achievability scheme for the second case is based on a random coding argument.
\end{abstract}

\IEEEpeerreviewmaketitle

\section{Introduction}
Most wireless networks are half-duplex constrained, i.e. the network nodes cannot transmit and receive simultaneously. In order to handle the half-duplex constraint, transmission protocols deterministically split the time of each network node into transmission and reception periods. This approach is easy to realize since nodes do not have to change rapidly between their transmission and reception modes. However, the approach is suboptimal from an information theoretic point of view. It does not take into account that the throughput of each half-duplex node can be increased by allowing it to choose the transmission-reception patterns in dependence of the information to be sent. 

This observation goes back to~\cite{kra04} that introduced a binary, deterministic channel model for half-duplex constrained relays and demonstrated, using the example of a three node line network, that larger rates as compared to time-sharing are possible by modulating the operation modes of the relay based on the underlying information \cite{kra07}. In \cite{VijWonLok07}, the capacity of the degraded half-duplex relay channel was derived. The authors also noted that the schedule of the relay has to carry information in order to achieve the capacity.

An extension of this result to line networks with multiple sources was presented in \cite{LuHaKo08} and \cite{LuHaKo09}. In what follows, we refer to the intermediate nodes, i.e., the second to second-last nodes in the cascade, as {\it relays} since they must relay the first node's message to the last node that is the destination for all messages. Within the setup of~\cite{LuHaKo09} a source and a subset of the relays deliver independent information to the destination under the assumption that adjacent node pairs are connected by noise-free $(q+1)$-ary pipes. A coding scheme based on timing was proposed, and based on the asymptotic behavior of the coding scheme, the capacity of deterministic relay cascades of arbitrary length and a single source was established. If the cascade includes a certain number of relays with their own information, the coding scheme achieves the cut-set bound provided that the rates of the relay sources fall below individual thresholds. 



In the present paper, we treat deterministic half-duplex line networks with two sources where either the first or the last relay in the cascade is the second source. In both cases, we establish the capacity region of rates at which both sources can transmit independent information to a common sink. 

If the first relay acts as a source, it is shown that the capacity region is the cut-set region. This improves a result derived in \cite{LuHaKo09} which says that the cut-set bound is achievable if the rate of the relay source falls below a certain threshold. In order to understand the new step in the achievability scheme, we briefly describe the scheme in \cite{LuHaKo09}. Therein, the source node encodes its information by means of transmission symbols and idle symbols. An idle symbol indicates a channel use without transmission. The relays encode received information with the transmission pattern and with the value of the transmission symbols. Based on this, codes can be constructed which allow the nodes to cooperate in a sense that each node controls the transmission pattern applied by the next node. Hence, new information injected by the relay source is not allowed to be represented by the transmission pattern since, otherwise, the previous node is not able to control the applied transmission pattern. This is, in fact, the reason why the rate of the relay sources cannot exceed a certain threshold. In the new scheme, the relays still use the original idea. However, the source regards its link to the relay source as an erasure channel  where the erasures are a consequence of the half-duplex constraint. Hence, the source and the relay source do not cooperate anymore which enables the relay source to represent own information with transmission patterns. It turns out that this new step is necessary to achieve all points in the cut-set region.

In the second part of this paper, we focus on the case where the last relay in the cascade acts as a source. The capacity region is derived by means of a random coding argument.

\section{Network Model}
\begin{figure}[t]
\psfrag{X0}{\small{\textrm{$X_0$}}}
\psfrag{Xi}{\small{\textrm{$X_k$}}}
\psfrag{X1}{\small{\textrm{$X_1$}}}
\psfrag{Yi}{\small{\textrm{$Y_k$}}}
\psfrag{Y1}{\small{\textrm{$Y_1$}}}
\psfrag{X_m}{\small{\textrm{$X_{m}$}}}
\psfrag{Y_m}{\small{\textrm{$Y_{m}$}}}
\psfrag{Relay-Source}{\footnotesize{\textrm{Relay Source}}}
\psfrag{i}{\small{\textrm{${k}$}}}
\psfrag{i-1}{\small{\textrm{${k-1}$}}}
\psfrag{1}{\scriptsize{$1$}}
\psfrag{2}{\scriptsize{$2$}}
\psfrag{Channel}{\footnotesize{\textrm{Channel}}}
\psfrag{ENC}{\footnotesize{\textrm{Enc.}}}
\psfrag{DEC}{\footnotesize{\textrm{Dec.}}}
\psfrag{MUX}{\footnotesize{\textrm{Mux}}}
\psfrag{W0}{\small{\textrm{$W_0$}}}
\psfrag{Wi-1}{\small{\textrm{$W_{k-1}$}}}
\psfrag{W1}{\small{\textrm{$W_1$}}}
\psfrag{Wm}{\small{\textrm{$W_m$}}}
\centering
\epsfig{file=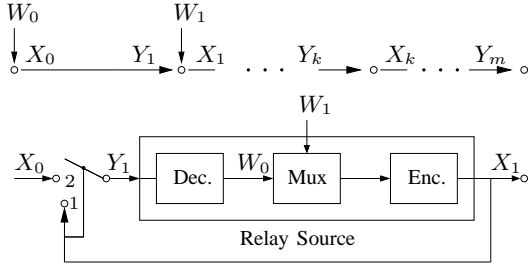, scale = 0.7}
\caption{A noiseless relay cascade with two sources. The link model is illustrated by means of feedback. If relay~$1$ is transmitting, the switch is in position~$1$ otherwise in position~$2$.}
\vspace{-0.5cm}
\label{fig:sys_model}
\end{figure}
Consider a discrete memoryless relay cascade as depicted in Fig.~\ref{fig:sys_model}. Each node is labeled by a distinct number from $\mathcal{V}=\{0,\dots,m\}$ with $m>0$. The integers $0$ and $m$ refer to the source and sink, respectively, while all remaining integers $1$ to $m-1$ represent half-duplex constrained relays, i.~e. relays which cannot transmit and receive at the same time. The connectivity within the network is described by the set of edges $\mathcal{E}=\{(k,k+1):0\leq k \leq m-1\}$, i.e. the ordered pair $(k,k+1)$ represents the communications link from node $k$ to node $k+1$. The output of the $k$th node, which is the input to channel $(k,k+1)$ is denoted as $X_k$ and takes values on the alphabet $\mathcal{X}_k=\mathcal{Q}_k\cup \{\textrm{N}\}$ where $\mathcal{Q}_k$ denotes the transmission alphabet of node~$k$ while the idle symbol ``N'' signifies a channel use in which node $k$ is not transmitting. The input of the $k$th node, which is the output of channel $(k-1,k)$ is denoted as $Y_k$ and is given by
\begin{align}
\label{relay_ch_model}
Y_{k}=\left\{
	\begin{array}{ll}
		X_{k-1}, & \mbox{if }X_{k}=\textrm{N}\\ 
		X_{k}, & \mbox{if }X_{k} \in \mathcal{Q}_k 
	\end{array}
	\right.\\
Y_{m}=X_{m-1}.\hspace{2.6cm}
\end{align}
where $1 \leq k \leq m-1$. Channel model (\ref{relay_ch_model}) captures the half-duplex constraint as follows. Assume relay $k$ is in its transmission mode, i.e. $X_k \in \mathcal{Q}_k$. Then relay~$k$ hears itself ($Y_k=X_k$) but cannot listen to relay $k-1$ or, equivalently, relay $k$ and relay $k-1$ are disconnected. However, if relay $k$ is not transmitting, i.~e. $X_k=\textrm{N}$, it is able to listen to relay~$k-1$ via a noise-free $|\mathcal{X}_{k-1}|$-ary pipe ($Y_k=X_{k-1}$). Another interpretation of the channel model is that the output $X_k$ of each relay $k$ controls the position of a switch which is placed at its input. If relay~$k$ is transmitting, the switch is in position~$1$ otherwise it is in position~$2$ (see Fig.~\ref{fig:sys_model}). Since a pair of nodes is either perfectly connected or disconnected, we obtain a deterministic network with $p(y_1,\dots,y_{m}|x_0,\dots,x_{m-1})\in\{0,1\}$.

At the beginning of a new block $b$ of $n$ channel uses, source node $0$ and relay $k \in \{1,m-1\}$ produce a uniformly and independently drawn message $W_{0,b} \in \left \{1,\dots,2^{nR_0}\right\}$ and $W_{k,b} \in \left \{1,\dots,2^{nR_k}\right\}$, respectively. Based on the received sequence in block $b$, sink node $m$ forms the estimates $\hat{w}_{0,b-(m-1)}$ and $\hat{w}_{k,b-(m-1-k)}$ of $W_{0,b-(m-1)}$ and $W_{k,b-(m-1-k)}$. We assume the following encoding functions
\begin{eqnarray}
x_{0i}&=&f_{0i}(W_0) \\
x_{ki}&=&f_{ki}(W_k,Y_k^{i-1}) \\
x_{li}&=&f_{li}(Y_l^{i-1}),\qquad \forall l \neq \{0,k,m\}. 
\end{eqnarray}
The first subscript describes the node number while the second subscript $i$ corresponds to the time instance where $1 \leq i \leq n$. Moreover, $Y_k^{i-1}$ is used as short hand notation for the set $\{Y_{k1},\dots,Y_{k,i-1}\}$. 
\section{The First Relay is a Source}
\begin{figure}[t]
\psfrag{(a)}{\footnotesize{(a)}}
\psfrag{(b)}{\footnotesize{(b)}}
\psfrag{(c)}{\footnotesize{(c)}}
\psfrag{R0}{\footnotesize{\textrm{$R_0$ (bits per use)}}}
\psfrag{R1}{\footnotesize{\textrm{$R_1$ (bits per use)}}}
\centering
\epsfig{file=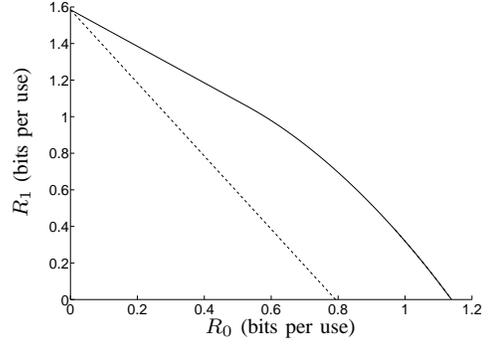, scale = 0.18}
\caption{Capacity region (\ref{Example_1}) is given by the solid curve. The time-sharing region is bounded by the dashed line.}
\label{Fig:Rate_region_two_sources_binary}
\vspace{-0.4cm}
\end{figure}

\begin{theorem}
The capacity region $\mathcal{C}$ of the line network of Fig.~\ref{fig:sys_model}, where node~$0$ and relay node~$1$ are sources, is
\begin{align}
   \set{C} = \bigcup \left\{
      \begin{array}{l}
         R_0 \le H(Y_1|X_1)\\
         R_0+R_1 \le H(Y_m)\\
         R_0+R_1 \le \min_{2\leq i\leq m-1} H(Y_i|X_i)
      \end{array}
      \right\} .
\label{C_special_case}
\end{align}
The union is over all probability distributions of the form 
\begin{equation}
P_{X_0}(\cdot) P_{X_1}(\cdot) P_{X_2|X_1}(\cdot) P_{X_3|X_2}(\cdot) \dots P_{X_{m-1}|X_{m-2}}(\cdot).
\label{Markov_Th1}
\end{equation}
\label{Theorem1}
\end{theorem}
\begin{proof}
We start with the achievability of $\set{C}$. At the end of block~$b-1$, node $0$ and relay node $1$ choose new messages $w_{0,b}$ and $w_{1,b}$, respectively, which are sent in block $b$ by means of the sequences $\ul{x}_0(w_{0,b})$ and $\ul{x}_1(w_{0,b-1},w_{1,b})$. The remaining relays $i$, $2\leq i \leq m-1$, forward older messages. In particular, relay $i$ sends $\ul{x}_i(w_{0,b-i},w_{1,b-(i-1)})$ in block $b$.

\textit{Coding: }
\begin{itemize}
\item At node $m-1$\cite{LuHaKo09}: Node $m-1$ represents information by taking $n_{m-1}<n$ transmission symbols per block of length $n$ from the alphabet $\set{Q}_{m-1}$ and by allocating the $n_{m-1}$ symbols to the transmission block. Thus, $|\set{Q}_{m-1}|^{n_{m-1}} {n\choose n_{m-1}}$ different sequences $\ul{x}_{m-1}\left(w_{0,b-(m-1)},w_{1,b-(m-2)}\right)$ are available at relay~$m-1$. Observe that $|\set{Q}_{m-1}|^{n_{m-1}}$ equals the number of possible distinct sequences when the $|\set{Q}_{m-1}|$-ary symbols are located at fixed slots while ${n \choose n_{m-1}}$ equals the number of possible transmission-listen patterns.  
\item At node $i$, $1\leq i \leq m-2$\cite{LuHaKo09}: For each transmission-listen pattern used by node $i+1$, node $i$ generates a codebook. For a particular pattern, node $i$ allocates $n_i$ transmission symbols from the alphabet $\set{Q}_i$ in all possible ways to the $n-n_{i+1}$ listen slots of the pattern. The slots of the pattern, in which node $i+1$ transmits, are filled with idle symbols ``N''. This procedure generates a certain number of transmission-listen patterns used by node $i$.
\item Due to the above codebook construction, adjacent nodes can cooperate since each node $i\geq 1$ knows the messages to be forwarded by the next node and, thus, is aware of the applied codeword. The construction guarantees that adjacent nodes $i$ and $i+1$, $i \geq 1$, do not transmit at the same time. 
\item At node $0$: In contrast to \cite{LuHaKo09}, node~$0$ does not adapt to the transmission-listen patterns used by node~$1$. Instead it uses an optimal point to point erasure channel code with alphabet $\set{X}_0$ for encoding $W_{0,b}$. Output symbols $Y_1$ of link $(0,1)$ are erased with a probability of $1-p_{X_1}(\textrm{N})$, i.e. the erasure probability is equal to the fraction of time in which node~$1$ transmits. It should be noted that node $0$ transmits a part of the information in the timing of the transmission symbols since the erasure code makes use of symbol ``N''.
\end{itemize}

\textit{Achievable Rates:}
The capacity of a $|\set{X}_0|$-ary erasure channel with erasure probability $1-p_{X_1}(\textrm{N})$ equals $p_{X_1}(\textrm{N})\log|\set{X}_0|$ achieved by a uniform input distribution over $\set{X}_0$. Due to the channel model, we clearly have
\begin{eqnarray}
H(Y_1|X_1)&=&H(X_{0}|X_1=N)\nonumber\\
&\leq& p_{X_1}(\textrm{N})\log|\set{X}_0|
\label{upper_bound_ach_R}
\end{eqnarray}
with equality if $p_{X_0|X_1}(\cdot|N)$ is the uniform distribution over $|\set{X}_0|$. Thus, an optimal erasure channel code for the link $(0,1)$ satisfies $R_0 = H(Y_1|X_1) - \epsilon$ with $\epsilon \rightarrow 0$ as $n\rightarrow \infty$.

Further, we know from the results derived in \cite[Sec. III.A]{LuHaKo09} that
\begin{equation}
|\set{Q}_i|^{n_i} {n-n_{i+1}\choose n_{i}} \rightarrow 2^{nH(Y_{i+1}|X_{i+1})}\qquad\textrm{as }n\rightarrow \infty.
\label{no_sequences}
\end{equation}
for $0\leq i \leq m-2$. For $i=m-1$, the exponent in (\ref{no_sequences}) becomes $H(Y_m)$ (with $n_m=0$). Hence, $R_0+R_1 \le \min_{2\leq i\leq m-1} H(Y_i|X_i)$ and $R_0+R_1 \le H(Y_m)$.

The converse is immediate since the bounds of (\ref{C_special_case}) correspond to the cut-set upper bound\cite[Sec. IV]{LuHaKo09}. (\ref{Markov_Th1}) follows from the following consideration. Again, due to the channel model
\begin{equation}
H(Y_i|X_i)=p_{X_i}(N)H(X_{i-1}|X_i=N),
\end{equation}
so that $H(Y_i|X_i)$ is a function of $p_{X_{i-1}X_i}(\cdot)$ for all $2 \le i \le m-1$. Hence, without restriction we may assume the Markov chain $X_1 - \dots - X_{m-1}$. Further, we can choose a uniform $p_{X_0|X_1}(\cdot|N)$ over $|\set{X}_0|$ since this achieves the upper bound (\ref{upper_bound_ach_R}). Clearly, such a distribution also exists when $X_0$ is independent of $X_1,\dots,X_{m-1}$. 
\end{proof}
\begin{remark}
Theorem~\ref{Theorem1} shows that the capacity region of the considered line network is equal to the cut-set region. This improves a result in \cite{LuHaKo09} which says that the cut-set bound is achieved when the rate of the relay source falls below a certain threshold. The new ingredient here is that the relay source is allowed to encode its own information in the timing of transmission symbols. In fact, node $0$ accepts that a part of its information is erased by node~$1$. We point out that this approach, namely to treat the link to the relay source~$k$ as an erasure channel, is not cut-set bound achieving if $k\geq 2$.
\end{remark}
\begin{example}
We apply Theorem~\ref{Theorem1} to a line network composed of three nodes where the first two nodes have their own information. The alphabets are $\set{X}_0=\set{X}_1=\{0,1,\textrm{N}\}$. This example has already appeared in \cite{LuHaKo08,LuHaKo09}. However, we are now able to characterize the complete capacity region. Moreover, the approach here is easier since we can restrict attention to independent $X_0$ and $X_1$. By choosing $P_{X_0}(\cdot)$ to be the uniform distribution and, further, by assigning the same probability masses to $X_1=0$ and $X_1=1$ (due to symmetry), we obtain the following expression for the capacity region
\begin{align}
   \set{C} = \bigcup \left\{
      \begin{array}{l}
         R_0 \le p_{X_1}(\textrm{N})\log 3\\
         R_0+R_1 \le (1-p_{X_1}(\textrm{N}))\log2 + h(p_{X_1}(\textrm{N})) \\
      \end{array}
      \right\}. 
\label{Example_1}
\end{align}
The union is over $p_{X_1}(\textrm{N})$ and $h(\cdot)$ denotes the binary entropy function. $\set{C}$ is depicted in Fig.~\ref{Fig:Rate_region_two_sources_binary}. Note that the region bounded by the dashed line contains the rates which are achievable when the time of the relay is deterministically split into transmission and reception periods. In order to obtain this region, time-sharing between $(R_0,R_1)=(0.5\log_23,0)$ and $(0,\log_23)$ bits per use has to be performed.
\end{example}
\section{The Last Relay is a Source}
In the remainder, we will make use of the following notation
\begin{eqnarray}
w_{0,b-[i+j;i+k]}&\stackrel{def}{=}& \{w_{0,b-(i+j)},\dots,w_{0,b-(i+k)}\} \nonumber \\
X_{[l;t]}&\stackrel{def}{=}& \{X_l,\dots,X_t\} \nonumber.
\end{eqnarray}

\begin{lemma}\cite[Th. 14.2.3]{cover-thomas:it-book}
Let $A_\epsilon^{(n)}$ denote the typical set for the probability mass function $p(x_1,\dots,x_n)$ and let
\begin{eqnarray}
P(\ul{X}_1=\ul{x}_1,\dots,\ul{X}_n=\ul{x}_n)=\prod_{l=1}^n p(x_{1l}|x_{3l},\dots,x_{nl})\nonumber\\
p(x_{2l}|x_{3l},\dots,x_{nl})p(x_{3l},\dots,x_{nl}). \nonumber
\end{eqnarray}
Then
\begin{equation}
P\{(\ul{X}_1,\dots,\ul{X}_n) \in A_\epsilon^{(n)}\}\doteq 2^{-n(I(X_1;X_2|X_3,\dots,X_n) \pm 6 \epsilon)}. \nonumber
\end{equation}
\label{Lemma1}
\end{lemma}
\begin{theorem}
The capacity region $\mathcal{C}$ of the line network of Fig.~\ref{fig:sys_model}, where node~$0$ and relay node~$m-1$ are sources, is
\begin{align}
   \set{C}=\bigcup \left\{
      \begin{array}{l}
         R_0 \le \min_{1\leq i \leq m-1} H\left(Y_i| X_{i}\right) \\
	 R_{m-1} \le H(Y_m|U) \\
         R_0+R_{m-1} \le H(Y_m)
      \end{array}
	\right\}.
\end{align}
The union is over all probability distributions of the form 
\begin{equation}
P_{X_0} P_{X_1} P_{X_2|X_1} P_{X_3|X_2} \dots P_{X_{m-2}|X_{m-3}} P_{U|X_{m-2}} P_{X_{m-1}|U}.
\label{Markov_Th2}
\end{equation}
\begin{remark}
$\set{C}$ is equal to the cut-set region if there exists a probability distribution for each boundary point such that $X_{m-1}$ is independent of $U$. Otherwise, $\set{C}$ is smaller than the cut-set region.
\end{remark}

\label{Theorem2}
\end{theorem}

\section{Proof Outline of Theorem \ref{Theorem2}}
\subsection{Achievability}
\textit{Random codebook generation:}
\begin{itemize}
\item Split $W_0$ into $B$ sub-blocks $W_{0,b}$, $b=1,2,\ldots,B$, that each take on $2^{nR_0}$ values. Similarly, split $W_{m-1}$ into $B$ sub-blocks $W_{m-1,b}$, $b=1,2,\ldots,B$, that each take on $2^{nR_{m-1}}$ values.
\item Node $m-1$ generates at random $2^{nR_0}$ independent sequences of length $n$, $\ul{u}\left(w_{0,b-(m-1)}\right)$, $w_{0,b-(m-1)} \in \{1,\dots,2^{nR_0}\}$, according to $\prod_{l=1}^n p(u_l)$. 
\item Codebook at node $m-1$: On each of the sequences $\ul{u}\left(w_{0,b-(m-1)}\right)$, node $m-1$ superposes a random codebook with $2^{nR_{m-1}}$ codewords $\ul{x}_{m-1}\left(w_{0,b-(m-1)},w_{m-1,b}\right)$ using  $\prod_{l=1}^n p(x_{m-1,l}|u_l)$.
\item Codebook at node $m-2$: For each $\ul{u}\left(w_{0,b-(m-1)}\right)$, node $m-2$ generates $2^{nR_0}$ independent sequences $\ul{x}_{m-2}\left(w_{0,b-[m-2;m-1]}\right)$ according to $\prod_{l=1}^n p(x_{m-2,l}|u_l)$.
\item Codebook at node $i$, $0 \leq i < m-2$: For each $\ul{x}_{i+1}\left(w_{0,b-[i+1;m-1]}\right)$ node $i$ generates at random $2^{nR_0}$ independent sequences $\ul{x}_{i}\left(w_{0,b-[i;m-1]}\right)$ according to $\prod_{l=1}^n p(x_{i,l}|x_{i+1,l})$.
\end{itemize}

\textit{Encoding: }At the beginning of each block $b$, node $i$, $0 \leq i \leq m-2$, has the estimates\footnote{The source knows its own messages. However, for simplicity, we will also denote this message with a hat. The same is done for the relay source~$m-1$.} $\hat{w}_{0,b-i-l}$ of $w_{0,b-i-l}$, $l \geq 0$. To send the estimate $\hat{w}_{0,b-i}$, node~$i$ selects the codeword $\ul{x}_{i}\left(\hat{w}_{0,b-[i;m-1]}\right)$.

Similarly, at the beginning of block $b$, node $m-1$  has the estimates $\left\{\hat{w}_{0,b-(m-1)-l},\hat{w}_{m-1,b-l}\right\}$ of $\left\{w_{0,b-(m-1)-l},w_{0,b-l}\right\}$, $l \geq 0$. To send the pair $\left\{\hat{w}_{0,b-(m-1)},\hat{w}_{m-1,b}\right\}$, node~$m-1$ selects the codeword
\begin{equation}
\ul{x}_{m-1}\left(\hat{w}_{0,b-(m-1)},\hat{w}_{m-1,b}\right). \nonumber
\end{equation}
Every node $i$, $1 \leq i \leq m$, receives the sequence $\ul{y}_i\left(b\right)$ in block $b$.

\textit{Decoding: }At the end of block $b$, sink node $m$ performs the following $\epsilon$-typicality check in order to determine $\hat{w}_{0,b-(m-1)}$ and $\hat{w}_{m-1,b}$:
\begin{eqnarray}
\left\{\ul{u}\left(\hat{w}_{0,b-(m-1)}\right),\ul{x}_{m-1}\left(\hat{w}_{0,b-(m-1)},\hat{w}_{m-1,b}\right),\ul{y}_{m}(b)\right\}\nonumber \\
\in A_\epsilon^{(n)}(U,X_{m-1},Y_m). 
\label{t_check_m}
\end{eqnarray}
By Lemma \ref{Lemma1}, it follows that the error probability of (\ref{t_check_m}) is smaller than
\begin{equation}
2^{-n(I(U,X_{m-1};Y_m)-6\epsilon)}.
\label{P_error_m-1}
\end{equation}
Further, if the estimate $\hat{w}_{0,b-(m-1)}$ is known at the sink, the error probability regarding the estimate $\hat{w}_{m-1,b}$ is smaller than 
\begin{equation}
2^{-n(I(X_{m-1};Y_m|U)-6\epsilon)}.
\label{P_error_m-1_U}
\end{equation}
Similarly, at the end of block $b$, node $i$, $1 \leq i \leq m-1$, performs the following $\epsilon$-typicality check in order to determine $\hat{w}_{0,b-i}$:
\begin{eqnarray}
\left\{\ul{x}_{i-1}\left(\hat{w}_{0,b-[i-1;m-1]}\right),\ul{x}_{i}\left(\hat{w}_{0,b-[i;m-1]}\right),\ul{y}_{i}(b)\right\} \nonumber \\
\in A_\epsilon^{(n)}(X_{i-1},X_i,Y_{i}).
\label{t_check}
\end{eqnarray}
According to Lemma \ref{Lemma1}, the error probability of (\ref{t_check}) regarding the estimate $\hat{w}_{0,b-(i-1)}$ is smaller than
\begin{equation}
2^{-n(I(X_{i-1};Y_i|X_{i})-6\epsilon)}.
\label{P_error_i}
\end{equation}
Now, by considering all possible error events we obtain from (\ref{P_error_m-1}), (\ref{P_error_m-1_U}) and (\ref{P_error_i}) that 
\begin{align}
  \set{R}=\bigcup\left\{
      \begin{array}{l}
         R_0 \le \min_{1\leq i \leq m-1} H\left(Y_i| X_i\right) \\
	 R_{m-1} \le H(Y_m|U) \\
         R_0+R_{m-1} \le H(Y_m)
      \end{array}
	\right\}
\end{align}
is an achievable region. Observe that the exponents of the error probabilities can be simplified since $Y_i$ is a function of $X_i, X_{i-1}$. 
\subsection{Converse}
Consider the following bounds, where $P_{b,0}$ and $P_{b,m-1}$ are the average bit error probabilities when decoding $W_0$ and $W_{m-1}$ at the destination node $m$. For $1 \leq l \leq m-1$, we have
\begin{align}
 & nR_0(1-h(P_{b,0})) \\
 & \stackrel{(a)}{\le} I\left(W_0;Y_m^n\right) \\
 & \le I\left(W_0 ; W_{m-1} Y_{l}^n Y_{m}^n\right) \\
 & \stackrel{(b)}{=} \sum_{i=1}^n I\left(W_0 ; Y_{li} Y_{mi} | W_{m-1} Y_{l}^{i-1} Y_{m}^{i-1}\right) \\
 & \stackrel{(c)}{=} \sum_{i=1}^n I\left(W_0 ; Y_{li} | W_{m-1} Y_{l}^{i-1} Y_{m}^{i-1}\right) \\
 & \stackrel{(d)}{=} \sum_{i=1}^n I\left(W_0 ; Y_{li} | W_{m-1} Y_{l}^{i-1} Y_{m}^{i-1} X_{l}^i\right) \\
 & \stackrel{(e)}{=} \sum_{i=1}^n H\left(Y_{li} | W_{m-1} Y_{l}^{i-1} Y_{m}^{i-1} X_{l}^i\right) \\
 & \stackrel{(f)}{\le} \sum_{i=1}^n H\left(Y_{li} | X_{li}\right) \\
 & \stackrel{(g)}{=} n H\left(Y_{l} | X_{l},Q\right) \\
 & \stackrel{(h)}{\le} n H\left(Y_l | X_l\right) 
\end{align}
where
\begin{itemize}
  \item (a) follows by Fano's inequality
  \item (b) follows from the chain rule for mutual information and from the independence of $W_0$ and $W_{m-1}$
  \item (c) follows by Markovity
  \item (d) follows because $X_l^i$ is a function of $Y_l^{i-1}$ for all $1\leq l < m-1$ and $X_{m-1}^i$ is a function of $Y_{m-1}^{i-1}$ and $W_{m-1}$
  \item (e) follows because $W_0$ determines $X_0^{i-1},\dots,X_{m-2}^{i-1}$ what, in turn, determines $Y_1^{i-1},\dots,Y_{m-2}^{i-1}$ and $Y_{m-2}^{i-1}$
  \item (f) conditioning does not increase entropy
  \item (g) follows by defining $Q$ to be a time-sharing random variable with $Y_l:=Y_{lQ}$, $X_l:=X_{lQ}$.
  \item (h) conditioning does not increase entropy
\end{itemize}

Further, we have the bounds
\begin{align}
 & nR_{m-1}(1-h(P_{b,{m-1}})) \\
 & \stackrel{(a)}{\le} I\left(W_{m-1};Y_{m}^n\right) 
\end{align}
\begin{align}
 & \le I(W_{m-1} ; W_0 Y_{m}^n) \\
 & \stackrel{(b)}{=} \sum_{i=1}^n I\left(W_{m-1} ; Y_{mi} | W_0 Y_m^{i-1}\right) \\
 & \stackrel{(c)}{=} \sum_{i=1}^n I\left(W_{m-1} ; Y_{mi} | W_0 Y_m^{i-1} X_{m-1}^{i-1}\right) \\
 & \stackrel{(d)}{=} \sum_{i=1}^n I\left(W_{m-1} ; Y_{mi} | W_0 Y_{[1;m]}^{i-1} X_{[0;m-1]}^{i-1}\right) \\
 & \stackrel{(e)}{=} \sum_{i=1}^n H\left(Y_{mi} |  W_0 Y_{[1;m]}^{i-1} X_{[0;m-1]}^{i-1}\right) \\
 &  \stackrel{(f)}{\le} \sum_{i=1}^n H\left(Y_{mi} |V_i\right) \\
 &  \stackrel{(g)}{=} nH\left(Y_{m} |U\right)
\end{align}
where
\begin{itemize}
  \item (a) follows by Fano's inequality
  \item (b) follows from the chain rule for mutual information and from the independence of $W_0$ and $W_{m-1}$
  \item (c) follows because $Y_m^{i-1}=X_{m-1}^{i-1}$ 
  \item (d) follows because $W_0$ determines $Y_1^{i-1},\dots,Y_{m-1}^{i-1}$
  \item (e) follows because $W_0$ and $W_{m-1}$ determine $Y_{m,i}$
  \item (f) follows by defining $V_i=\left (X_{m-1}^{i-1},Y_{m-1}^{i-1}\right)$ and from the fact that conditioning does not increase entropy
  \item (g) follows by defining $Q$ to be a time-sharing random variable with $U:=(V_Q,Q)$ and $Y_m:=Y_{mQ}$.
\end{itemize}

Concerning the sum-rate, we obtain
\begin{align}
 & nR_0(1-h(P_{b,0})) + nR_{m-1}(1-h(P_{b,m-1})) \\
 & \stackrel{(a)}{\le} I(W_0;Y_m^n) + I(W_{m-1};Y_m^n) \\
 & \le I(W_0;Y_m^n) +  I(W_{m-1}; W_0 Y_m^n) \\
 & \stackrel{(b)}{=} I(W_0 W_{m-1} ; Y_m^n ) \\
 & = \sum_{i=1}^n I(W_0 W_{m-1} ; Y_{mi} | Y_m^{i-1}) \\
 & \stackrel{(c)}{\le} \sum_{i=1}^n H(Y_{mi} ) \\
 & \stackrel{(d)}{=} nH(Y_m|Q) \\
 & \stackrel{(e)}{\le} nH(Y_m).
\end{align}
where
\begin{itemize}
  \item (a) follows by Fano's inequality
  \item (b) follows from the independence of $W_0$ and $W_{m-1}$
  \item (c) follows since $W_0$ and $W_{m-1}$ determine $Y_{m,i}$ and from the fact that conditioning does not increase entropy
  \item (d) follows by defining $Q$ to be a time-sharing random variable and $Y_m:=Y_{mQ}$
  \item (e) conditioning does not increase entropy.
\end{itemize}

It remains to check (\ref{Markov_Th2}). Observe that $X_{m-1,i}$ is a function of $Y_{m-1}^{i-1}$ and $W_2$. Since $X_{0i},\dots,X_{m-2,i}$ do not depend on $W_2$, we have the Markov chain $X_{0i},\dots,X_{m-2,i} - X_{m-1}^{i-1}Y_{m-1}^{i-1} - X_{m-1,i}$. Hence, we have
\begin{align}
  & P(u,x_0,\dots,x_{m-1}) \\
  & = P(u)P(x_0,\dots,x_{m-1}|u) \\
  & = P(u)P(x_{0i},\dots,x_{m-2,i}|i,x_{m-1}^{i-1},y_{m-1}^{i-1}) \\
  & \hspace{1cm}\cdot P(x_{m-1,i}|i,x_{m-1}^{i-1},y_{m-1}^{i-1}) \nonumber \\
  & = P(u)P(x_0,\dots,x_{m-2}|u) P(x_{m-1}|u)
\end{align}
which shows that
\begin{align}
   X_0,\dots,X_{m-2}-U-X_{m-1}.
\end{align}
Finally, by the explanations in the last section of the proof of Theorem~\ref{Theorem1} we have the Markov chain
\begin{align}
   X_1 - \dots - X_{m-2}-U-X_{m-1}
\end{align}
and the independence of $X_0$ from $X_1,\dots,X_{m-1},U$.
\qed
\section{Discussion}
An obvious extension is to allow any relay in the cascade to act as second source. However, the solution for this case turns out to be elusive. Though developing achievable rate regions using superposition random coding is straightforward, proving a converse seems to be more difficult. An intuitive explanation is that having the second source located at the first or the last link offers greater freedom for choosing a coding strategy as compared to the other links. This is related to the fact that the first source does not receive information while the sink node does not send information and, therefore, both nodes are not affected by the half-duplex constraint. One could also think of extending the erasure coding technique outlined in the proof of Theorem~\ref{Theorem1}. In particular, if all nodes before the relay source use independent erasure codes, the relay source would be able to send own information in the timing of transmission symbols. However, it can be shown that this approach does not achieve the cut-set bound and, therefore, a converse is missing again.
\section*{Acknowledgment}
\label{sec:ack}
T. Lutz and C. Hausl were supported by the European Commission in the framework of the FP7 (contract n. 215252). G. Kramer was supported by NSF Grant CCF-09-05235.
\bibliographystyle{unsrt}
\bibliography{ISIT_2010_v9}

\end{document}